\documentclass[prl,twocolumn,aps,superscriptaddress]{revtex4}

\usepackage{graphicx}
\usepackage{amssymb,latexsym,amsthm,amsmath}    
\usepackage{color}
\usepackage{graphicx}

\begin{document}

\title{Explosive transitions to synchronization in networked phase oscillators}

\author{I. Leyva}
\affiliation{Complex Systems Group, Universidad Rey Juan Carlos, 28933 M\'ostoles, Madrid, Spain}
\affiliation{Center for Biomedical Technology, Universidad Polit\'ecnica de Madrid, 28223 Pozuelo de Alarc\'on, Madrid, Spain}
\author{I. Sendi\~na-Nadal}
\affiliation{Complex Systems Group, Universidad  Rey Juan Carlos, 28933 M\'ostoles, Madrid, Spain}
\affiliation{Center for Biomedical Technology, Universidad Polit\'ecnica de Madrid, 28223 Pozuelo de Alarc\'on, Madrid, Spain}
\author{J. Almendral}
\affiliation{Center for Biomedical Technology, Universidad  Polit\'ecnica de Madrid, 28223 Pozuelo de Alarc\'on, Madrid, Spain}
\author{A. Navas}
\affiliation{Center for Biomedical Technology, Universidad  Polit\'ecnica de Madrid, 28223 Pozuelo de Alarc\'on, Madrid, Spain}
\author{M. Zanin}
\affiliation{Center for Biomedical Technology, Universidad  Polit\'ecnica de Madrid, 28223 Pozuelo de Alarc\'on, Madrid, Spain}
\affiliation{Faculdade de Ci\^encias e Tecnologia, Departamento de Engenharia Electrot\'ecnica, Universidade Nova de Lisboa, Portugal}
\affiliation{Innaxis Foundation \& Research Institute, Jos\'e Ortega y Gasset 20, 28006, Madrid, Spain.}
\author{D. Papo}
\affiliation{Center for Biomedical Technology, Universidad
  Polit\'ecnica de Madrid, 28223 Pozuelo de Alarc\'on, Madrid, Spain}
\author{J.M. Buld\'u}
\affiliation{Complex Systems Group, Universidad Rey Juan Carlos, 28933 M\'ostoles, Madrid, Spain}
\affiliation{Center for Biomedical Technology, Universidad Polit\'ecnica de Madrid, 28223 Pozuelo de Alarc\'on, Madrid, Spain}
\author{S. Boccaletti}
\affiliation{Center for Biomedical Technology, Universidad  Polit\'ecnica de Madrid, 28223 Pozuelo de Alarc\'on, Madrid, Spain}

\begin{abstract}
We introduce a condition for an ensemble of networked phase oscillators to
feature an abrupt, first-order phase transition from an unsynchronized to a synchronized state.
This condition is met in a very wide spectrum of situations, and for various oscillators' initial
frequency distributions. We show that the occurrence of such transitions is always accompanied by the spontaneous
emergence of frequency-degree correlations in random network architectures. We also discuss
ways to relax the condition, and to further extend the possibility for the first-order transition
to occur, and illustrate how to engineer {\it magnetic-like} states of synchronization. Our findings thus
indicate how to search for abrupt transitions in real-world applications.

\end{abstract}

\maketitle

Critical phenomena in complex networks, and the emergence of abrupt dynamical transitions 
in the macroscopic state of a system are currently a subject of the utmost interest \cite{kar07,doro08,rad09,grass11}. 
As far as the synchronization of an ensemble of networked phase 
oscillators is concerned \cite{physrepsincro,physrepnetworks,physreparenas}, the occurrence of a
first-order phase transition was described, so far, in two rather special and apparently opposite situations: 
{\it i)} an all-to-all coupling architecture, with oscillators' 
frequencies evenly spaced, that is, a measure-zero realization of a uniform distribution \cite{Pazo05}, and {\it ii)} a scale-free connection topology,  
where a positive correlation between the heterogeneity of the connections 
and the frequencies of the oscillators is introduced {\it ad hoc} \cite{Gardenes11,Leyva12}.

In this Letter, we show how a sharp, discontinuous phase transition can occur, 
instead, as a {\it generic feature} of the synchronization of networked phase oscillators. 
Precisely, we initially give a condition for the transition from unsynchronized to synchronized states to be first-order, and
demonstrate how such a condition is easy to attain for any
oscillators' initial frequency distribution. 
We then show how such transitions are always accompanied by the
spontaneous emergence of frequency-degree correlation features. 
Third, we show that the considered condition can be even softened in
several cases. Finally, we illustrate, as a possible application, the option of expressing
{\it magnetic-like} states of synchronization with the use of such
transitions. 

We consider a network of Kuramoto \cite{kuramoto} oscillators:
\begin{equation}
  \frac{d\phi_i}{dt}=\omega_i +   d\sum_{i=1}^N a_{ij}\sin(\phi_j-\phi_i),
  \label{eq:kuramoto}
\end{equation}
\noindent where $\phi_i$ is the phase of the $i^{th}$ oscillator ($i=1,...,N$), 
$\omega_i$ is its associated natural frequency drawn from a generic 
frequency distribution $p(\omega)$, $d$ is the coupling strength, and 
$\{ a_{ij} \}$ are the elements of the adjacency matrix that uniquely defines the graph. 
The classical order parameter for system (\ref{eq:kuramoto}) is 
$r(t)= \frac{1}{N} \mid \sum_{l=1}^N e^{i \phi_l(t)} \mid$, 
and the level of phase synchronization can be monitored by looking 
at the value of $S=\langle r(t)\rangle_T$, where $\langle \dots\rangle_T$ denotes a time average 
with $T\gg 1$. 
Furthermore, for each oscillator $i$, we denote by ${\cal N}(i)$ the set of oscillators linked to it.

As the coupling strength $d$ increases, system (\ref{eq:kuramoto}) 
undergoes a phase transition from the unsynchronized ($S \sim 1/\sqrt{N}$) to a 
synchronous ($S \sim 1$) state, where all oscillators ultimately acquire
the same frequency. 
For this phase transition to display a first-order feature, we have to avoid that
any oscillator behaves as the core of a clustering process,
where its neighbors begin to aggregate to the synchronous
state smoothly \cite{clustering}. This could be realized by considering
a certain frequency difference with its neighbors.
We have observed that a condition in order to achieve a first order phase transition is:

\begin{itemize}
\item[(A)] for each oscillator $i$, all nodes $j$ belonging to ${\cal N}(i)$ 
satisfy $|  {\omega_i}-{\omega_j}| > \gamma_c$
\end{itemize}

 \begin{figure}[t]
 \includegraphics[width=0.5\textwidth]{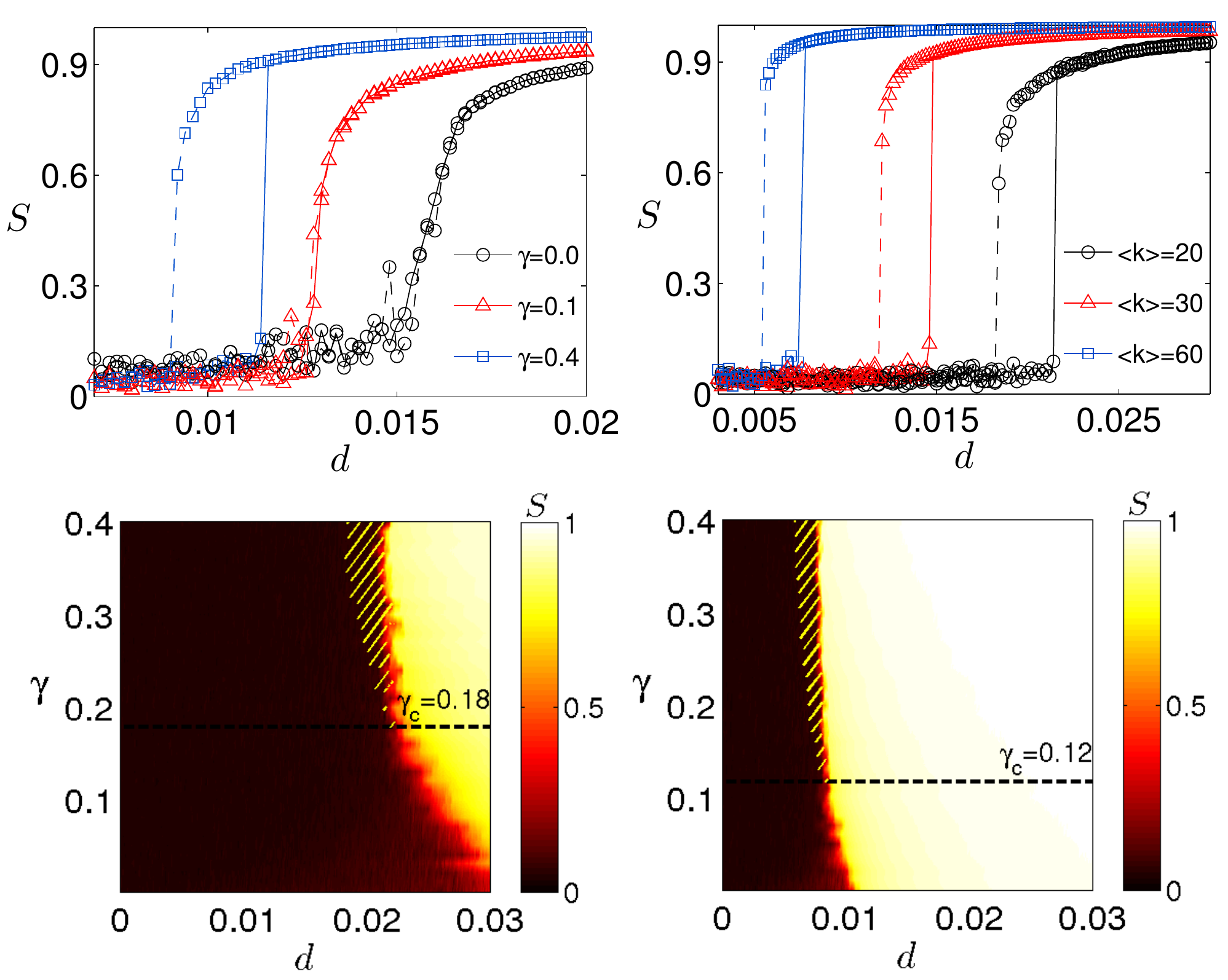}
 \caption{(Color online) (Top panels) Phase synchronization level $S$
   (see text for definition) {\it vs.} the coupling strength $d$,
   (left panel) for different values of the gap $\gamma$ and $\langle
   k\rangle =40$, and (right panel) different values of the average degree  $\langle k\rangle$ with $\gamma=0.4$.
The frequencies of the oscillators are here drawn from a uniform distribution
within the range $[0,1]$. See text for the construction procedure of the networks.
In both panels, the continuous (dashed) lines refer to the forward (backward) simulations.
(Bottom panels)  $S$ in the parameter space ($d,\gamma$). The values are color coded,
according to the color bars. (Left panel) $\langle k \rangle=20$ and
(right panel) $\langle k \rangle =60$. The horizontal dashed lines mark the
separation between the region of the parameter space where a
second-order transition occurs (below the line) and that in which the
transition is instead of the first order type (above the line). The
striped area delimits the hysteresis region.
 \label{fig1} }
 \end{figure}

 \begin{figure}[t]
 \includegraphics[width=0.5\textwidth]{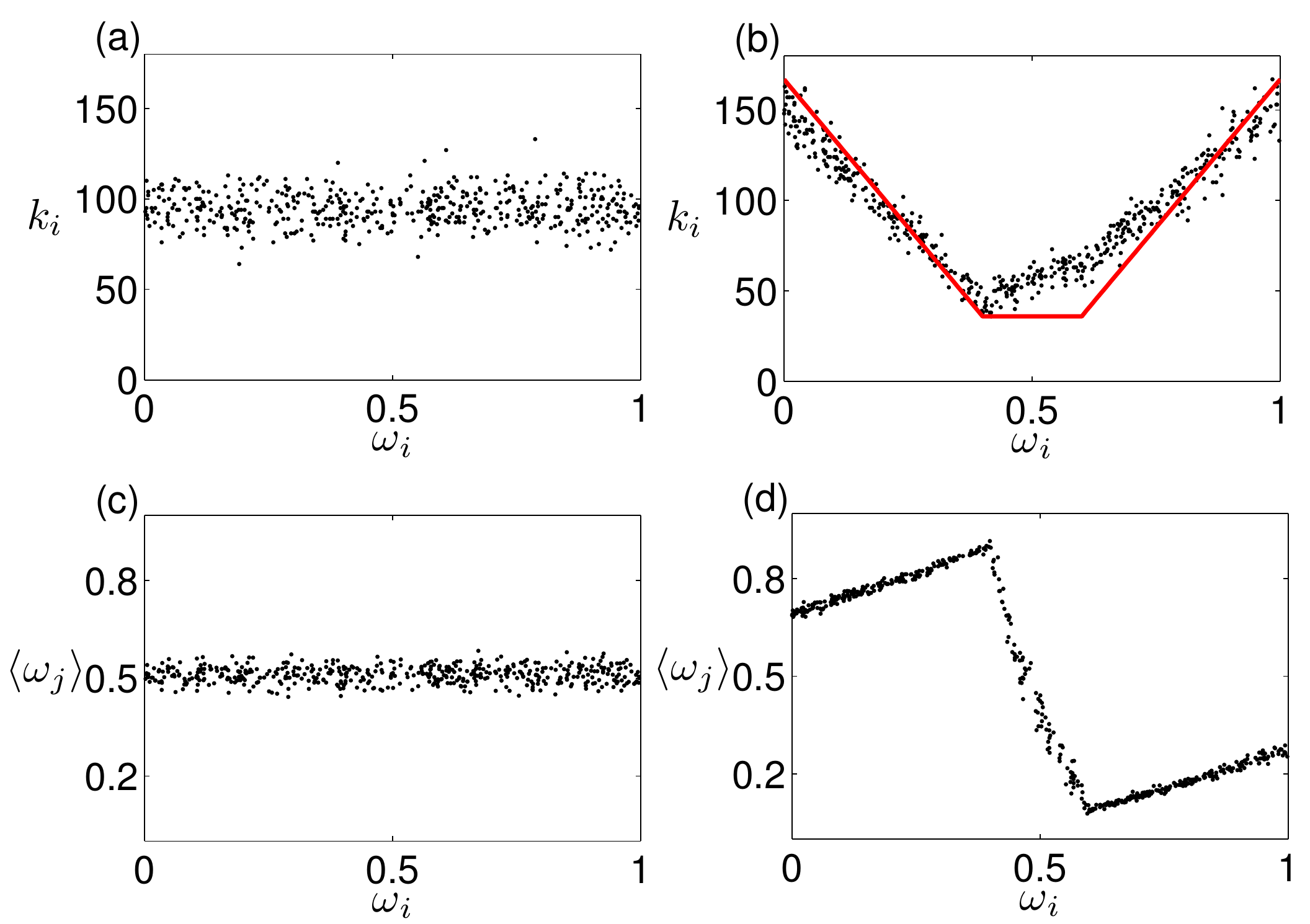}
 \caption{(Color online) (Top panels) Degree $k_i$ that each node achieves
  after the network construction is completed {\it vs.} its natural
  frequency  $\omega_i$. $\langle k \rangle=100$ and
  frequency gaps (a) $\gamma=0.0$, and  (b) $\gamma=0.4$. 
  The red solid line in panel (b) represents the theoretical prediction
  $f(\omega)$ (see text). (Bottom panels) Average of the local natural frequency $\langle \omega_j\rangle$ for $j\in
  {\cal N}(i)$ {\it vs.} the natural frequency $\omega_i$ of the
  $i^{th}$ node of the network for (c) $\gamma=0.0$ and (d) $\gamma=0.4$.}
 \label{fig2}
 \end{figure}

\noindent Condition (A) is tantamount to imposing a minimal value for 
the frequency difference between linked nodes.
We now fix $N=500$, and illustrate the synchronization route for several frequency
distributions $p(\omega)$ when condition (A) is met. To this purpose, we consider network topologies
resulting from the following procedure: {\it i)} we assign natural frequencies $\{ \omega_i \}$, 
drawn from a distribution $p(\omega)$,  to the $N$ oscillators; 
{\it ii)} we randomly pick a pair $(i,j)$ of oscillators, and form a link between 
them only if the value of $| {\omega_i}-{\omega_j}|$ exceeds a given gap $\gamma$; 
{\it iii)} we repeat point {\it ii)} until
the desired number of links $L$ in the graph is formed. 
After a final check on the connectedness of the resulting network, the procedure 
yields Erd\"os-Renyi-like \cite{erdos} 
topologies with an average degree 
$\langle k\rangle \equiv \frac{L}{2N}$. 
We then use the obtained adjacency matrix to simulate system (\ref{eq:kuramoto}), and monitor 
the state of the network as a function of the coupling strength $d$, by gradually increasing 
the value of $d$ in steps $\delta d= 10^{-4}$, from $d=0$.
At each step, a long transient is discarded before the 
data are acquired for further processing. Moreover, insofar as we are looking for a first-order 
phase transition (and thus for an expected associated synchronization hysteresis),  
simulations are also performed in the reverse way, i.e.\ starting from a given $d_{\max}$ 
(where the ensemble is phase synchronized), and gradually decreasing the coupling by 
$\delta d$ at each step. In what follows, the two sets of numerical trials are termed as 
{\em forward} and {\em backward}, respectively.

In Fig. \ref{fig1} we report the results obtained by setting $p(\omega)$ as a
uniform frequency 
distribution in the interval $[0,1]$. 
The top panels depict $S$ as a function of $d$. Namely, the top left (right) panel illustrates 
the case of a fixed mean degree $\langle k\rangle =40$ (of a fixed frequency gap $\gamma=0.4$), 
and reports the results of the forward and backward simulations at different values 
of $\gamma$ ($\langle k\rangle$). The bottom panels 
show $S$ in the parameter space $d-\gamma$ for $\langle k \rangle=20$ and
$\langle k \rangle =60$. 

Several pieces of information can be extracted from the Figure. 
First, the rise of a first order phase transition is a generic feature in the parameter space: 
the horizontal dashed lines in the bottom panels mark the value of $\gamma_{c}$, separating the two 
regions where a second-order transition (below the line) and a first-order transition (above the line) occurs.
Second, while increasing $\langle k\rangle$ of the network 
facilitates the occurrence of the explosive transition, as 
both the values of $\gamma$ and $d$ for which the first-order phase transition takes place shrink, 
it also reduces the region of hysteresis. 
This is consistent with the results of Ref. \cite{Pazo05} for an
all-to-all connected case, where in the limit $N \to \infty$ a
first-order phase transition has been predicted in the absence of hysteresis. 

Another relevant result is the spontaneous emergence of degree-frequency correlation 
features associated to the passage from a second- to a first-order phase transition. 
While such a correlation was imposed {\it ad hoc} in Refs. \cite{Gardenes11,Leyva12}, 
here condition (A) creates for each oscillator $i$ a 
frequency range around $\omega_i$, where links are forbidden. The final degree 
$k_i$ is proportional to the total probability for that oscillator to receive 
connections from other oscillators in the network, and therefore
to $1-\int_{\omega_i-\gamma}^{\omega_i+\gamma} p(\omega') d \omega'$. 
This is shown in Fig.~\ref{fig2}(a)-(b), where the degree $k_i$ that each node achieved after the network 
construction is completed is reported as a function of its natural frequency  
$\omega_i$, for $\langle k \rangle=100$. 
Panel (a) refers to the case $\gamma=0$ in which no degree-frequency 
correlation is present in the resulting network. In panel (b), instead, we report the case $\gamma=0.4$ 
(a value for which a first-order phase transition occurs) and the (conveniently normalized) 
function $f(\omega)=1-\int_{\omega-\gamma}^{\omega+\gamma} p(\omega') d \omega'$,
with $p(\omega)=1$ for $\omega \in [0,1]$, and $p(\omega)=0$ elsewhere, 
which gives evidence of the emergence of a very pronounced V-shape relationship 
between the frequency and the degree of the network's
nodes. Inspecting the average frequency of each oscillator's neighbors, we
also observe that condition (A) leads to the emergence of a bipartite-like network where low
frequency oscillators are mainly coupled to high frequency
oscillators, as shown in panels (c) and (d) for increasing  $\gamma$. 

 \begin{figure}
\centering 
\includegraphics[width=0.44\textwidth]{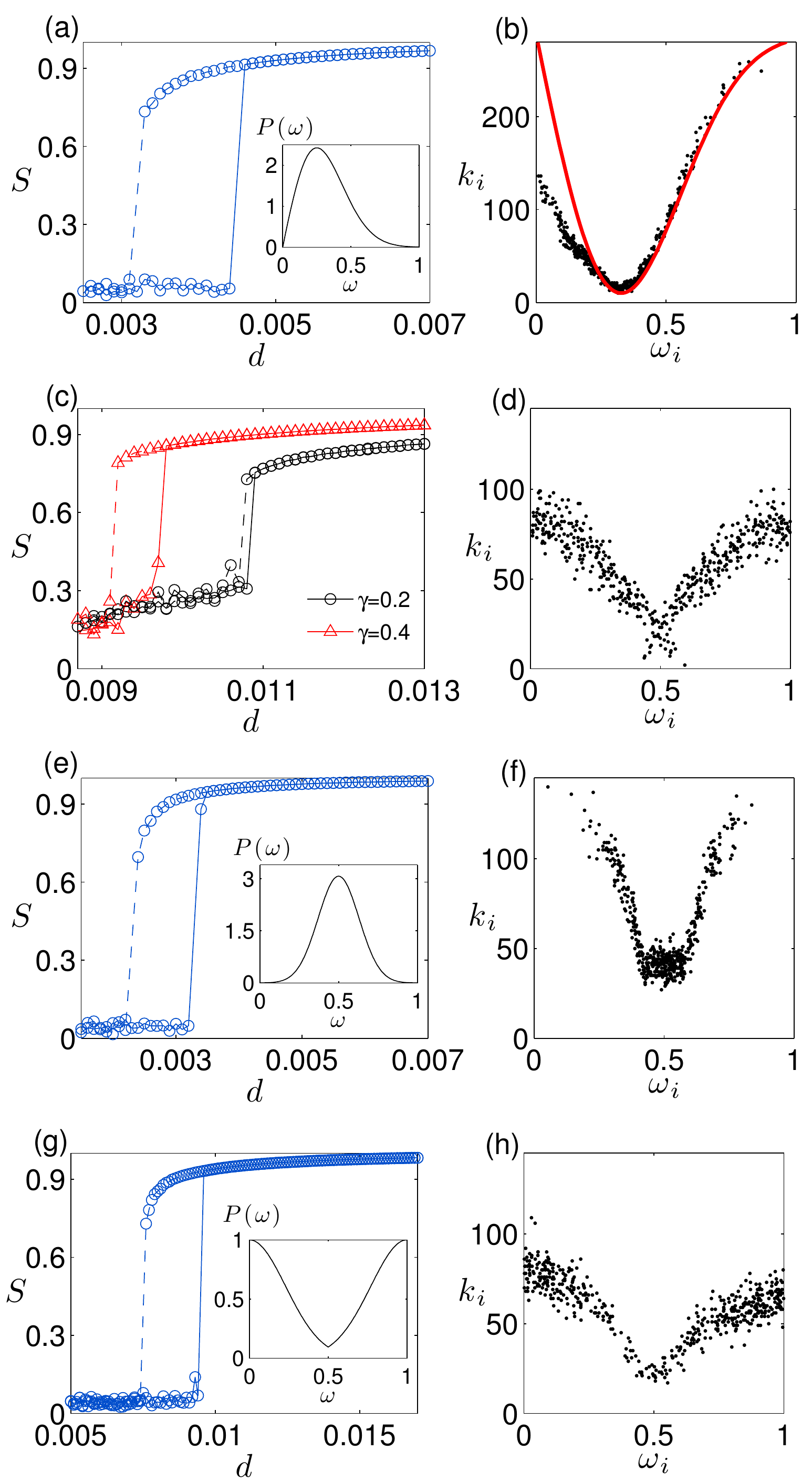}
\caption{(Color online) $S$ {\it vs.} $d$ resulting from the forward (continuous lines) and backward 
(dashed lines) simulations of system (\ref{eq:kuramoto}), 
displaying first-order phase transitions for different initial
  frequency distributions $p(\omega)$, defined on the interval
  $[0,1]$ (left panels), and distribution of the final node degree 
  $k_i$ {\it vs.} the corresponding oscillator's natural
frequency (right panels).   From top to bottom: (a)-(b) Rayleigh distribution
  for $\gamma=0.3$. In (b), the red solid line represents the
  theoretical prediction $f(\omega)$; (c)-(d) uniform distribution  using the local mean
  field condition for two values of $\gamma$. In panel (d)
  $\gamma=0.4$; 
 (e)-(f) Gaussian distribution with $Z=0.7$; (g)-(h) bimodal
  Gaussian with $Z=0.8$. The insets in panels (a), (e), and (g) report
  the  corresponding distribution $p(\omega)$. See text for the details on the specific construction 
  procedure used in each case. In all cases, $\langle k\rangle=60$.}
\label{fig3}
\end{figure}

We have verified that fulfillment of condition (A) leads to a first-order transition for a very wide class of
distributions of the oscillators' natural frequencies.
Figure~\ref{fig3}(a)-(b) shows, for instance, the case of the asymmetrical Rayleigh distribution,
conveniently re-scaled to the interval $[0,1]$, given by 
$p(\omega)=\frac{\omega}{\sigma^2} e^{-\frac{\omega^2}{2\sigma^2}}$
(see inset Fig.~\ref{fig3}(a) with $\sigma=0.25$). The results highlight the presence of the abrupt transition [Fig.\ref{fig3}(a)],
together with the emergence of a clear frequency-degree correlation,
well described, again, by the function $f(\omega)$ [Fig.\ref{fig3}(b)].

We now move to discussing several ways to soften condition (A), while still keeping the abrupt character of the
transition. For the first extension, we consider again the case of a uniform frequency distribution 
in the interval $[0,1]$. In this case, condition (A) can be relaxed as follows:

\begin{itemize}
\item[(A')] for each oscillator $i$, all nodes $j$ belonging to ${\cal N}(i)$ satisfy
$| {\omega_i}-\langle \omega_j\rangle | > \gamma_{c}$, where $\langle \dots \rangle$ 
indicates the average value over the ensemble ${\cal N}(i)$. 
\end{itemize}
The new condition is tantamount to softening condition (A) to the {\it local mean field} of
frequency differences in the neighborhood of each network node. Figure \ref{fig3}(c)-(d) reports the results
for networks obtained with a modified construction procedure, in which
pairs of randomly selected nodes are now linked if the value of $|{\omega_i}-\langle\omega_j\rangle |$ (averaged over
the set of nodes $j$ already linked to node $i$, and the one candidate to be further linked), exceeds a
 gap $\gamma$.
Again, an explosive transition occurs [Fig.~\ref{fig3}(c)] in correspondence to the emergence of frequency-degree correlations
[Fig.~\ref{fig3}(d)].

Furthermore, it is worth noticing that a strict application of condition (A) for non uniform frequency distributions implies
that oscillators at different frequencies would in general have a different number of available neighbors in the network. 
That's why a natural extension of condition (A) is to consider a frequency-dependent gap $\gamma (\omega)$ defined by
$\int_{\omega-\gamma}^{\omega+\gamma} p(\omega') d\omega' =Z$.
Panels (e)-(h) of Fig.~\ref{fig3} report the case of two symmetrical frequency distributions,
limited to the same frequency range $[0,1]$, for the sake of comparison:
{\it i)} a Gaussian distribution centered at $\omega=0.5$, and given by
$p(\omega)=\frac{1}{\sigma\sqrt{2\pi}}
e^{-\frac{(\omega-0.5)^2}{2\sigma^2}}$, with $\sigma=0.13$
[Fig.~\ref{fig3}(e)-(f)],
and {\it ii)} a bimodal distribution derived from the same Gaussian,
and given by $p(\omega)=\frac{1}{\sigma\sqrt{2\pi}} e^{-\frac{\omega^2}{2\sigma^2}}$ if $\omega<0.5$, and $p(\omega)=
\frac{1}{\sigma\sqrt{2\pi}} e^{-\frac{(\omega-1.0)^2}{2\sigma^2}}$
otherwise  [with $\sigma=0.23$, Fig.~\ref{fig3}(g)-(h)].
The gap condition for the construction of the network is now 
to fix the value of $Z$, and accept the pairing of nodes
when $|{\omega_i}-{\omega_j} | > \frac{1}{2}[\gamma(\omega_i)+\gamma(\omega_j)]$. 
Once again, an explosive transition is obtained, with pronounced
frequency-degree correlation features, as long as $p(\omega)$ is
symmetrical. 

\begin{figure}
  \centering
  \includegraphics[width=0.49\textwidth]{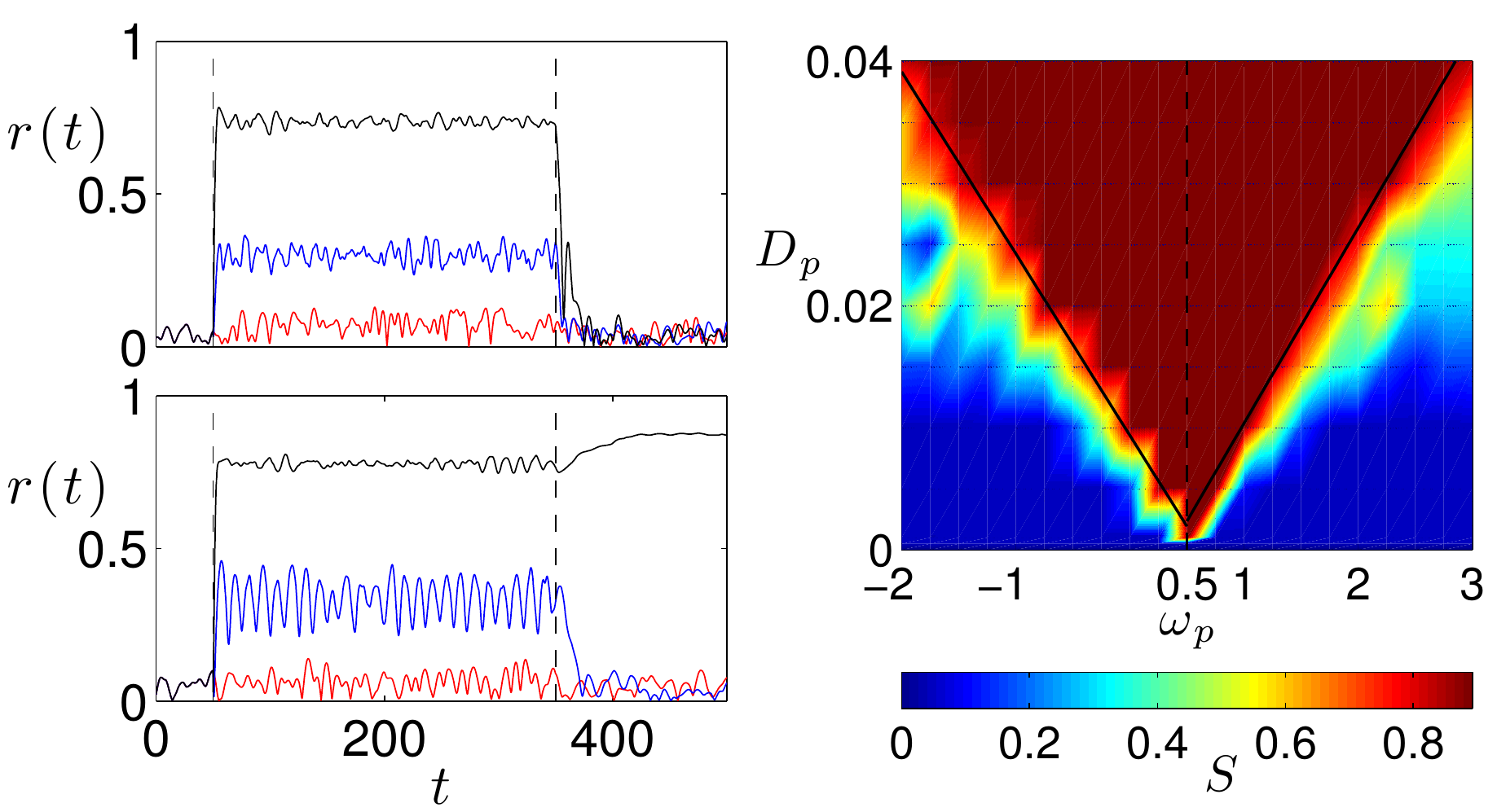}
  \caption{(Color online) (Left panels) Time evolution of the parameter $r(t)$
  (see text for definition) for $d=0.004$ (top panel,  outside the hysteresis region),
and $d=0.009$ (bottom panel, inside the hysteresis region).
The initial frequencies of the oscillators are taken from a uniform
distribution in the interval $[0,1]$. $\gamma=0.49$ and $\langle k\rangle=40$. 
$\omega_p=1.0$ and $D_p=0.0005$ (bottom red
    line), $D_p=0.005$ (middle blue line), and $D_p=0.02$ (top black line).
    The pacemaker is active from $t=50$ to $t=350$, as marked by
    the vertical dashed lines. (Right panel) Colormap of $S=\langle r(t)\rangle_{t>350}$ (coded as indicated in the color bar), 
    showing the region of the parameter space $D_p$-$\omega_p$
    where the {\it magnetic-like} state of synchronization
    is maintained after removal of the pacemaker. 
    \label{fig4}}
\end{figure}

Finally, we show how our findings allow expressing {\it magnetic-like} states of synchronization in a 
such ensemble of networked oscillators, provided that the coupling strength is set 
inside the hysteresis region of the first-order phase transition.
For this purpose, we consider again the case of an initial uniform distribution of the
oscillators' frequencies, and modify system (\ref{eq:kuramoto}) as follows: $\frac{d\phi_i}{dt}=\omega_i + 
D_p \sin( \phi_p - \phi_i) + d\sum_{i=1}^N a_{ij}\sin(\phi_j-\phi_i)$, where $D_p$ is the strength of 
a unidirectional connection to an external pacemaker (equal for all oscillators in the network),
and $\phi_p$ is the phase of the pacemaker obeying $\frac{d\phi_p}{dt}=\omega_p$.
Initially, the system freely evolves in the unsynchronized regime, with $D_p = 0$.
The pacemaker is then switched on, and $D_p$ is selected so that all oscillators are entrained
to the pacemaker phase. At a subsequent time, the pacemaker is switched off again, and
the state of the system is monitored. The left panels of Fig.~\ref{fig4} show that setting $d$ outside (inside)
the hysteresis region produces a final state that relaxes to the original 
unsynchronized behavior (that stays permanently in a synchronized configuration for sufficiently
large $D_p$ values). The right panel of the same Figure depicts the regions of the parameter space 
$D_p - \omega_p$ for which these {\it magnetic-like} states can ultimately be produced.

In conclusion, we provided a condition for the occurrence of abrupt phase transitions in networks 
of phase oscillators, proved its validity in several cases, and discussed several extensions. 
Our study generalizes previous results, and extends the possibility of encountering first-order phase transitions 
to a large variety of network topologies, as well as to a large
variety of frequency distribution of the oscillators.   
This indicates practical methods for engineering networks able to display critical phenomena, 
and the emergence of dynamical abrupt transitions in their macroscopic states. 
Furthermore, the evidence for the emergence of frequency-degree correlations in connection with these abrupt transitions,
may shed light on the mechanisms underlying the relationship between topology and dynamics 
in many real-world systems.

 Work supported by Ministerio de Educaci\'on y Ciencia, Spain, through grants FIS2009-07072 
 and from the BBVA-Foundation within the Isaac-Peral program of Chairs. Authors acknowledge also 
 the R\&D Program MODELICO-CM [S2009ESP-1691], and the usage of the resources, technical 
 expertise and assistance provided by supercomputing facility CRESCO
 of ENEA in Portici (Italy).

\end{document}